\documentclass[10pt,conference,compsocconf]{IEEEtran}
\usepackage[cmex10]{amsmath}
\pdfoutput=1
\usepackage{graphicx}
\usepackage[caption=false,font=footnotesize]{subfig}  %

\usepackage{tabularx}
\usepackage{threeparttable}

\usepackage{cite}
\usepackage{url}
\usepackage{color}

\begin{document}

\IEEEoverridecommandlockouts

\title{A Study of Bandwidth-Perception Management Mechanisms
\\in IEEE 802.16 Networks$^\star$\thanks{$^\star$An abridged version of this paper appeared in IEEE LCN 2009 \cite{andres2009}.}}
\author{
\IEEEauthorblockN{Andr\'es Arcia-Moret\IEEEauthorrefmark{1}, Yubo Yang\IEEEauthorrefmark{2}, Nicolas Montavont\IEEEauthorrefmark{2} and David Ros\IEEEauthorrefmark{2}}
\IEEEauthorblockA{\IEEEauthorrefmark{1}Universidad de Los Andes, M\'erida, Venezuela}
\IEEEauthorblockA{\IEEEauthorrefmark{2}Institut T\'el\'ecom / T\'el\'ecom Bretagne,  %
Rue de la Ch\^ataigneraie, CS 17607, 35576 Cesson S\'evign\'e cedex, France}
\IEEEauthorblockA{Email: amoret@ula.ve, \{yubo.yang, nicolas.montavont, david.ros\}@telecom-bretagne.eu}}

\maketitle

\begin{abstract}
Bandwidth request-grant mechanisms are used in 802.16 networks to manage the uplink bandwidth needs of subscriber stations (SSs). Requests may be sent by SSs to the base station (BS) by means of several mechanisms defined in the standard. Based on the incoming requests, the BS (which handles most of the bandwidth scheduling in the system) schedules the transmission of uplink traffic, by assigning transmission opportunities to the SSs in an implementation-dependent manner. In this paper we present a study of some bandwidth allocation issues, arising from the management of the \emph{perception} of subscriber stations' bandwidth needs at the base station. We illustrate how the bandwidth perception varies depending on the policy used to handle requests and grants. By means of ns-2 simulations, we evaluate the potential impact of such policies on the system's aggregate throughput when the traffic is composed of Best-Effort TCP flows.
\end{abstract}


\section{Introduction and Background}
\label{sec-introduction}

The IEEE 802.16 wireless metropolitan network standard \cite{802.16-2004,802.16e}, also known as WiMAX, defines high-performance mechanisms that provide last-mile, high-speed Internet services. Broadband wireless access for Metropolitan Area Networks is considered, in some scenarios, as a potential substitute for wired access technologies, and may also provide high-speed wireless connectivity to nomadic and mobile users.

In an IEEE 802.16 system, a base station (BS) allocates network resources to subscriber stations (SS). The IEEE 802.16 standards allow for dynamic on-demand (frame by frame) reservation of bandwidth for the uplink. To support various types of traffic (e.g., real-time traffic versus non-real time traffic), the IEEE 802.16 standard defines five Classes of Service (CoS), namely: Unsolicited Grant Service (UGS), extended real-time Polling Service (ertPS), real-time Polling Service (rtPS), non real-time Polling Service (ntrPS) and Best Effort (BE). These CoS allow setting various priorities to user traffic, which condition the rate, the delay and the jitter experienced by users' data flows. The prioritization is implemented at the MAC layer via a classifier, a scheduler and an admission control subsystem. %

\subsection{Uplink bandwidth allocation in IEEE 802.16 networks}
\label{sec-uplink-bw-alloc}

In 802.16 networks, bandwidth request-grant mechanisms are responsible of managing and satisfying the uplink bandwidth needs of subscriber stations. These mechanisms take care of the following functions: (1) allowing SSs to dynamically indicate their bandwidth requirements (which may vary over time); (2) managing the perception the BS has of SS bandwidth needs; (3) fulfilling those needs by granting uplink bandwidth allocations.

\subsubsection{Requesting for bandwidth}
\label{sec-bw-reqs}

An SS informs the BS of its bandwidth needs by sending a Bandwidth-Request Message (BW-REQ). Such message signals an amount of bytes waiting for transmission at an SS queue. The standard supports different methods by which an SS can send the BW-REQ to the BS.

In the \emph{polling-based} method, the BS polls the SSs for knowing their bandwidth needs. Basically, the polling mechanism can be triggered in two ways: either by the expiration of a timer in the BS, or a by an explicit request done by the SS (the ``poll-me'' bit in the grant management subheader of UGS connections). Polling can be done in a unicast manner, i.e., by allocating to a single SS enough bandwidth for sending a BW-REQ in an uplink burst. Also, the BS may poll groups of SSs (multicast polling) or even all the SSs (broadcast polling).

\begin{figure}[htbp]
\centering
\includegraphics[width=0.95\columnwidth]{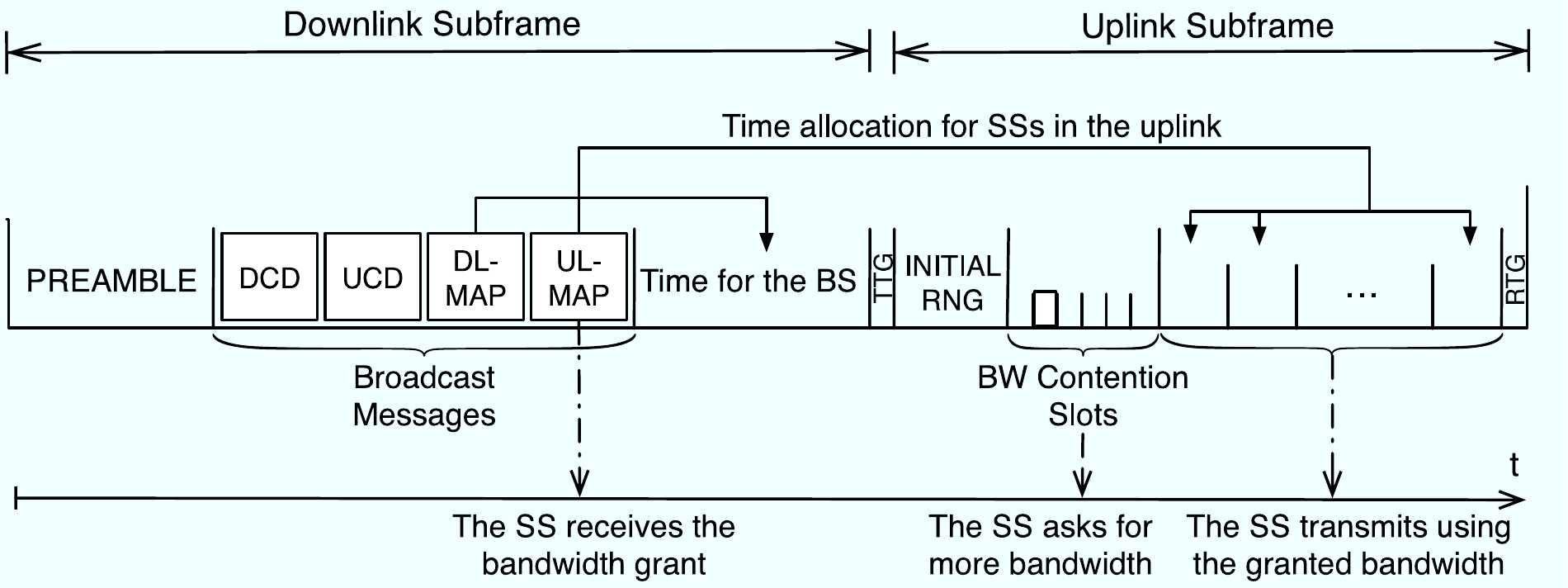}
\caption{IEEE 802.16 frame structure (TDD mode).}
\label{fig-wimax-frame}
\end{figure}

Multicast and broadcast polling are used for enabling \emph{contention-based} requests. 
In this method, the SSs contend to send BW-REQ messages using physical slots in a specific part of the uplink subframe, known as the \emph{bandwidth-request contention slots} (Fig.\ \ref{fig-wimax-frame}). Since the BW-REQs are not acknowledged, the SS considers that a request is lost if no data grant has been given within a certain time interval, called T16 in the 802.16-2004 standard\footnote{In the 802.16e amendment, T16 has been replaced by a so-called ``contention-based reservation timeout'', which corresponds to a number of consecutive UL-MAPs that do not contain a data grant.}. For instance, the loss could be caused by a collision (i.e., two or more SSs use the same contention slot to transmit a BW-REQ), or because the BS decides not to grant the requested bandwidth. Request losses are handled by a truncated binary exponential backoff (BEB) algorithm which allows to retransmit lost BW-REQs.

Finally, the \emph{piggybacking method} uses the grant management subheader to attach a bandwidth request to an uplink data packet.

\subsubsection{Managing bandwidth perception}
\label{sec-mgmt-bw-perception}

In order to be able to satisfy the bandwidth needs of the SSs, the BS has to keep track of incoming and outstanding requests---that is, it needs to have a ``perception'' of the status of the transmission queues at every SS. This information may be stored in an allocation table \cite{freitag07} (denominated traffic management table in \cite{cho05}). We will call \emph{Bandwidth Perception Manager} (BPM) the algorithm used for updating the allocation table, according to the collected SS queue size information and the granted requests. As we shall see in Section~\ref{sec-bw-policies}, different policies could be used to manage bandwidth perception; such policies are left unspecified in the standard.

\subsubsection{Satisfying SSs bandwidth needs}
\label{sec-scheduling}

The BS fulfills the SS bandwidth needs by allocating physical slots in the uplink subframe. Such allocations, defined in the UL-MAP, are handled by the BS uplink scheduler. The particular scheduling algorithm used is implementation-dependent. The standard states that bandwidth requests should be done on a per-connection basis, allowing the BS to properly implement a fair and QoS-oriented uplink scheduler \cite{eklund06}. Thus, each request must be associated with a connection using the connection identifier (CID). Note however that grants are made by SS, and not by connection. As explained in Section~\ref{sec-bw-policies}, in some cases this may lead to a mismatch between the allocation table at the BS and the actual needs of SSs.

\subsection{Goal and structure of the paper}
\label{sec-struct}

Most Internet traffic is carried by the Transmission Control Protocol (TCP). In WiMAX, the nrtPS and BE classes are intended for most TCP-based applications. BE traffic is given the remaining bandwidth after scheduling higher priority classes, without any guarantee on a minimum bandwidth, delay or jitter \cite{cicconetti06,chen05b}. Hence, BE is best suited for non-interactive (bulk) data traffic such as e-mail, though this CoS could be used for offering generic Internet access without any quality-of-service guarantees. %

In this paper, we present an analysis and simulation study of several bandwidth-perception management policies for 802.16 networks. Given that TCP accounts for most of the traffic in IP networks, the focus of the simulation study is on the potential interactions between such policies and the performance of TCP. The bandwidth perception may be computed through different policies, and is influenced by several mechanisms, such as the BS scheduler, the SS scheduler, or the way bandwidth perception is updated after a bandwidth allocation. As we will show later, the BS and the SS may be easily de-synchronized about the SS bandwidth needs. Since the bandwidth perception at the BS determines the scheduling, the difference between such perception and the actual needs of an SS may have a great impact on performance.

In addition, a new policy variant, which we call \emph{Decrease at Data Arrival with delayed BW-REQ handling} (DDA-d), is presented and compared with three other methods for handling bandwidth requests at the BS. Since \mbox{DDA-d} gives an adequate matching between the actual SS's bandwidth needs and its perception at the BS, we achieve full utilization of the uplink bandwidth, even at low levels of load.

This paper expands our previous work in \cite{andres2009} in two respects. First, we extend our prior analysis to consider two additional perception-management schemes. Second, we study the performance of such schemes under different conditions, in terms of BS queue sizes and of wireless loss rates.

The remainder of the paper is organized as follows. Section~\ref{sec-related} reviews the related work. Section~\ref{sec-bw-policies} describes several bandwidth perception management policies that may be used by a BS, and discusses some of their issues. A simulation study of such policies is presented in Section~\ref{sec-perfeval}. Finally, Section~\ref{sec-conclusions} concludes the paper.

\section{Related work}
\label{sec-related}

In IEEE 802.16 networks, the bandwidth scheduling is centralized at the BS. For the downlink, the BS schedules data bursts according to the data waiting in BS queues for downloading traffic. For the uplink traffic, data is scheduled after the BS processes BW-REQs sent by SSs. As already discussed in Section \ref{sec-uplink-bw-alloc}, BW-REQs are sent to the BS in several ways. In the sequel, we will consider only the multicast/broadcast polling method, which induces SSs to send their BW-REQs during contention periods. In order to improve BE traffic performance, researchers have focused on some of the critical subsystems of the bandwidth request-grant mechanism, namely: the length or the repartition of the contention period, and the sending frequency of BW-REQs. In the following, we briefly survey the consequent impact of these subsystems on the uplink bandwidth usage.

Cho et al.\ \cite{cho05} propose a general QoS architecture considering the different CoS. They show that maximum throughput for a number $N$ of uploading BE connections is reached when the contention window size (in number of slots) is equal to $N$. Similarly, Ni et al. \cite{Ni:2007fj} focus their study on the mean number of frames required to successfully send BW-REQs using contention or unicast polling. For a fixed contention window size, they show that periodical contention (i.e., for multicast or broadcast polling) is more efficient than unicast polling for a low BW-REQ arrival rate. On the other hand, for a larger number of SSs, multicast/broadcast polling is more efficient when the arrival rate of BW-REQ increases. Moreover, when the truncated-BEB parameters are well chosen, the delay (in average number of frames) it takes to successfully send a BW-REQ in contention mode is bounded. 

Vinel et al.\ \cite{gtc2g0i:vinel.a;zhang.y06a} show the benefits of grouping all CID bandwidth needs by SS to reduce the total number of BW-REQs, and to improve the mean delay to successfully convey a BW-REQ to the BS. Recently, Delicado et al.\  \cite{ts:delicado.j;delicado.f08a} have also found that such optimization 
improves the uplink throughput, at the expense of detailed information on the individual connections. Another way of treating BW-REQs is proposed in \cite{ce2i02itis:kobliakov.v;turlikov.a06a}. The authors propose to divide the contention period in two non-overlapping subsets. The first period is used to send new BW-REQs, and the second period to solve collisions. This method allows to decrease the mean delay time to successfully transmit BW-REQs, compared to the legacy BEB.

To reduce the number of collisions, Delicado et al. \cite{iwcncw2:delicado.j;ni.q09a} propose to independently adapt the contention windows (CW) of the SSs. As an extension to the standard, each connection sends its BW-REQs in an adapted (fairly constant) number of frames once the SS finds the adequate size of the CW (i.e., by exponentially increasing the CW). As a result, the different uplink BW-REQs experience less collisions. Moreover, after a silence period of a flow, a connection uses an adapted (reduced) back-off period based on the last size of the CW in which a BW-REQ was transmitted with success.  

A different solution to the uplink bandwidth management is proposed by Mojdeh et al. \cite{Mojdeh:2005lr}. As the bandwidth request-grant mechanism intrinsically produces an undesirable jitter and delay (e.g., due to BW-REQ losses and the resulting expiring of the timer T16), they propose a forecasting method based on data-mining to improve the performance of uplink real-time traffic. However, as Mojdeh et al.\ remark, there is also a risk of overestimating the bandwidth needs, thereby producing an underutilization of the uplink channel (by not fully using the allocated bandwidth).

The works described before have focused mostly on improving either the bandwidth usage for requests, or the number of collisions. To the best of our knowledge, bandwidth perception issues have been studied explicitly only by Cicconetti et al.\  \cite{cicconetti09}. They 
propose a mechanism called Bandwidth Request Reiteration ($BR^2$) that breaks eventual deadlocks on uplink bandwidth assignation, due to the loss of the requests during contention. Different from our study, in which the perception-synchronization algorithms are located only at the BS, the $BR^2$ mechanism requires the SSs to periodically send updates to the BS, informing about their bandwidth needs. These updating BW-REQs are directly sent after the BS has given bandwidth by contention, so $BR^2$ first ``hooks'' an SS to the BS and then it allows the SS to send requests without contention. This helps in maintaining the bandwidth needs updated at the BS.

\section{Bandwidth perception management in \\ 802.16 networks}
\label{sec-bw-policies}

The reception of BW-REQs allows the BS to modify its perception of the SSs needs for bandwidth. The standard defines two types of BW-REQ that an SS can use, namely, incremental and aggregate requests. 
An aggregate request tells the BS the current state (i.e., the total bandwidth needs) of an SS's connection queue, whereas an incremental request allows an SS to ask for more bandwidth for a given CID \cite{eklund06}. Therefore, after receiving a BW-REQ, the BS will add the bandwidth request size (BRS) in bytes to its current perception of bandwidth needs (incremental type), or reset its perception with the BRS value (aggregate type), depending on the request type \cite{802.16-2004}.

The default behavior of an SS is to issue incremental requests when new bytes arrive in a connection queue. Aggregate requests are necessary to ensure the self-correcting nature of the request-grant mechanism. Indeed, the perception at the BS can get out of sync with respect to the actual SSs needs, for reasons like unrecoverable PHY errors (e.g., BW-REQ losses), or because of decisions taken by the SS or the BS---e.g., the BS scheduler does not grant the desired bandwidth, or the SS uses the granted bandwidth for a different CID \cite{eklund06}. For these reasons, a timer indicates when an aggregate request should be sent instead of an incremental request. Yet the BS and an SS may get out of sync in many cases, depending on the scheduler policy (both at the BS and the SS), the bandwidth-perception management policy, and the timing in which the requests are made. We show later in this section that a request issued just after receiving a grant for a previous request may desynchronize the BS perception of the SS needs. Remark also that \emph{the SS does not know what particular policy is used by the BS to handle requests}, which may lead to misinterpretation at both the BS and the SS.

In order to show how an SS and its BS may be out of sync, we consider four types of bandwidth-perception management policies, following two main principles in updating the bandwidth perception:

\begin{itemize}
\item \textbf{Update based on the granted bandwidth:} \emph{Reset Per Grant} (RPG) policy and \emph{Decrease Per Grant} (DPG) policy.

\item \textbf{Update based on the actual uplink bandwidth usage:} \emph{Decrease at Data Arrival with Immediate BW-REQ handling} (DDA-i) policy and \emph{Decrease at Data Arrival with delayed BW-REQ handling} (DDA-d) policy.
\end{itemize}

In what follows we discuss the issues associated to these policies.

\subsection{Reset Per Grant (RPG)}
\label{sec-rpg}

The RPG policy is a simplistic approach to bandwidth perception management. This policy allows us to illustrate some subtle issues regarding both bandwidth perception at the BS \emph{and} the way the SS requests for bandwidth, since the two are intertwined.

With RPG, after granting bandwidth to an SS, the BS resets to zero the bandwidth perception for the served SS in the allocation table, even if not all of the requested bandwidth was granted. %

\begin{figure}[tbp]
\centering
\includegraphics[width=0.95\columnwidth]{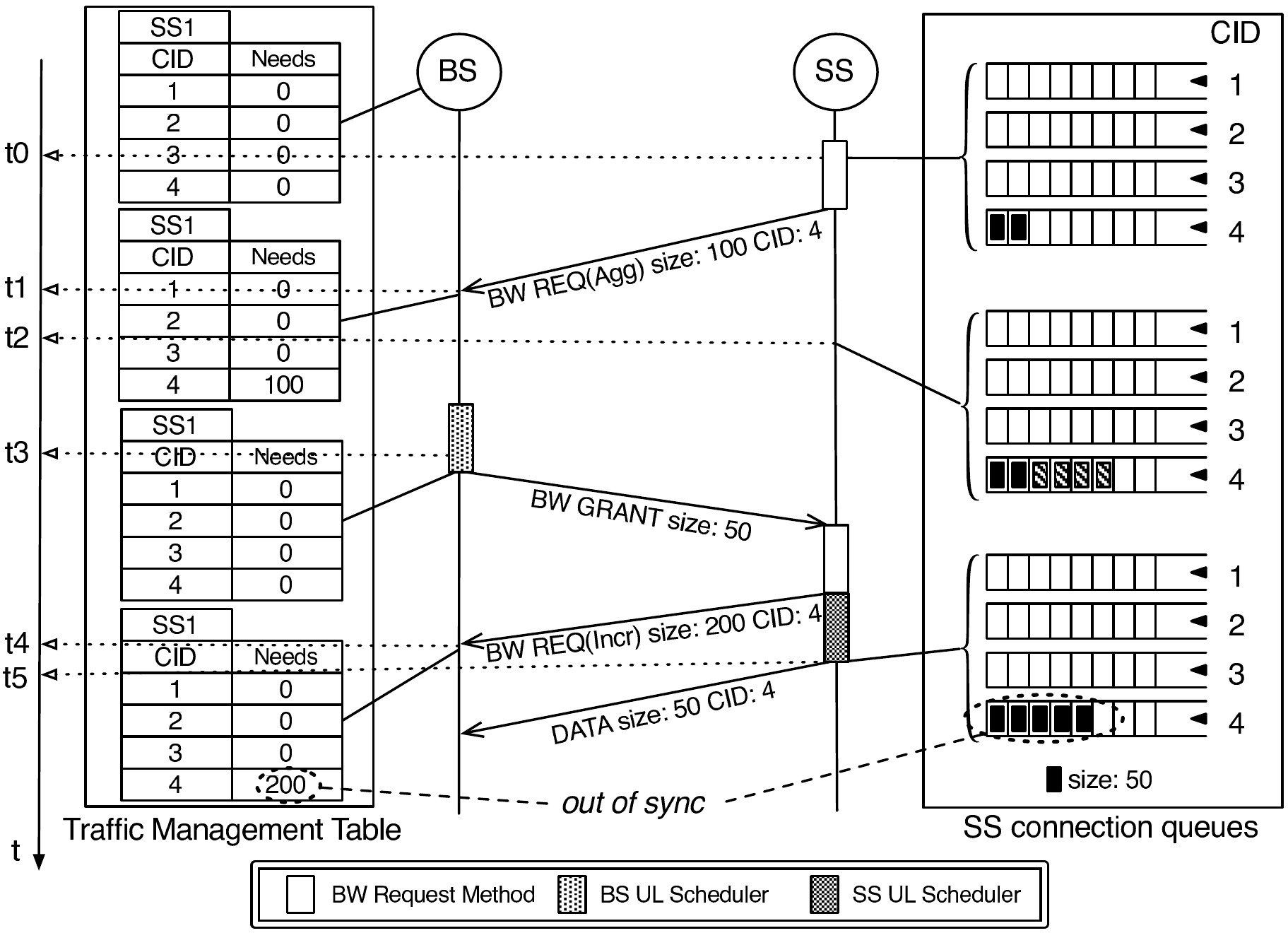}
\caption{RPG policy: the BS gets out of sync.}
\label{fig-rpg-out-sync}
\end{figure}

The obvious drawback of this policy is that it forgets the ungranted bandwidth needs, thus the BS perception of the SS queue status deviates from the actual situation. Figure~\ref{fig-rpg-out-sync} shows an example of this problem. At $t = t3$ the BS grants 50~bytes to the SS (i.e., half of what the SS asked for in an aggregate request, at $t = t0$), and resets the corresponding entry in the allocation table. However, at $t = t2$ an extra 200~bytes arrive at the CID~4 queue. Hence, the incremental request arriving at the BS at $t = t4$ demands 200~bytes, whereas (after the SS uses up the grant, at $t = t5$) there are 250~bytes in the queue.

The IEEE 802.16 standard being quite vague on the bandwidth perception management, the SS does not know the particular policy used by the BS to handle requests.
Hence, the SS cannot tell whether the BS will keep track of unfulfilled grants, so as to satisfy them later.
 So, if the SS had asked for 250~bytes (in an incremental request) instead of 200, it could well be that the BS, using a perception mechanism \emph{different} from RPG, had kept track of the outstanding 50~bytes from the first request---the BS would then ``see'' a 300-byte queue at the SS. On the other hand, if the new 250-byte request were sent in an aggregate BW-REQ, then such request could ``cross'' a new grant for the outstanding 50~bytes---analogously, the SS would end up having been granted a total of 350~bytes instead of 300.

Moreover, if SSs must re-request for non-granted bandwidth, then the total number of BW-REQs will increase. In the case of contention-based requests, this means the collision probability would increase, resulting in a higher delay for getting the corresponding grants, as shown in \cite{ts:delicado.j;delicado.f08a} and \cite{Ni:2007fj}.

\subsection{Decrease Per Grant (DPG)}
\label{sec-dpg}

With DPG, the BS updates the bandwidth perception when the BS uplink scheduler assigns bandwidth to an SS (i.e., when the UL-MAP is generated), by decrementing one or more entries in the table by the allocated amount.

Nonetheless, due to the fact that bandwidth requests are made on a per-connection basis and the grants are made per SS, the BS does not know for which connection it should decrease the bandwidth perception; it is the SS scheduler who determines how the allocation will be used. Thus, there exist different ways of implementing this policy. 

For instance, the BPM could opt for decreasing the bandwidth perception starting with the higher-QoS connection, then proceeding with lower-QoS connections, until the total amount of granted bandwidth is subtracted from the allocation table. This method is based on the assumption that the SS will serve higher-priority connections first.

\begin{figure}[tbp]
\centering
\includegraphics[width=0.95\columnwidth]{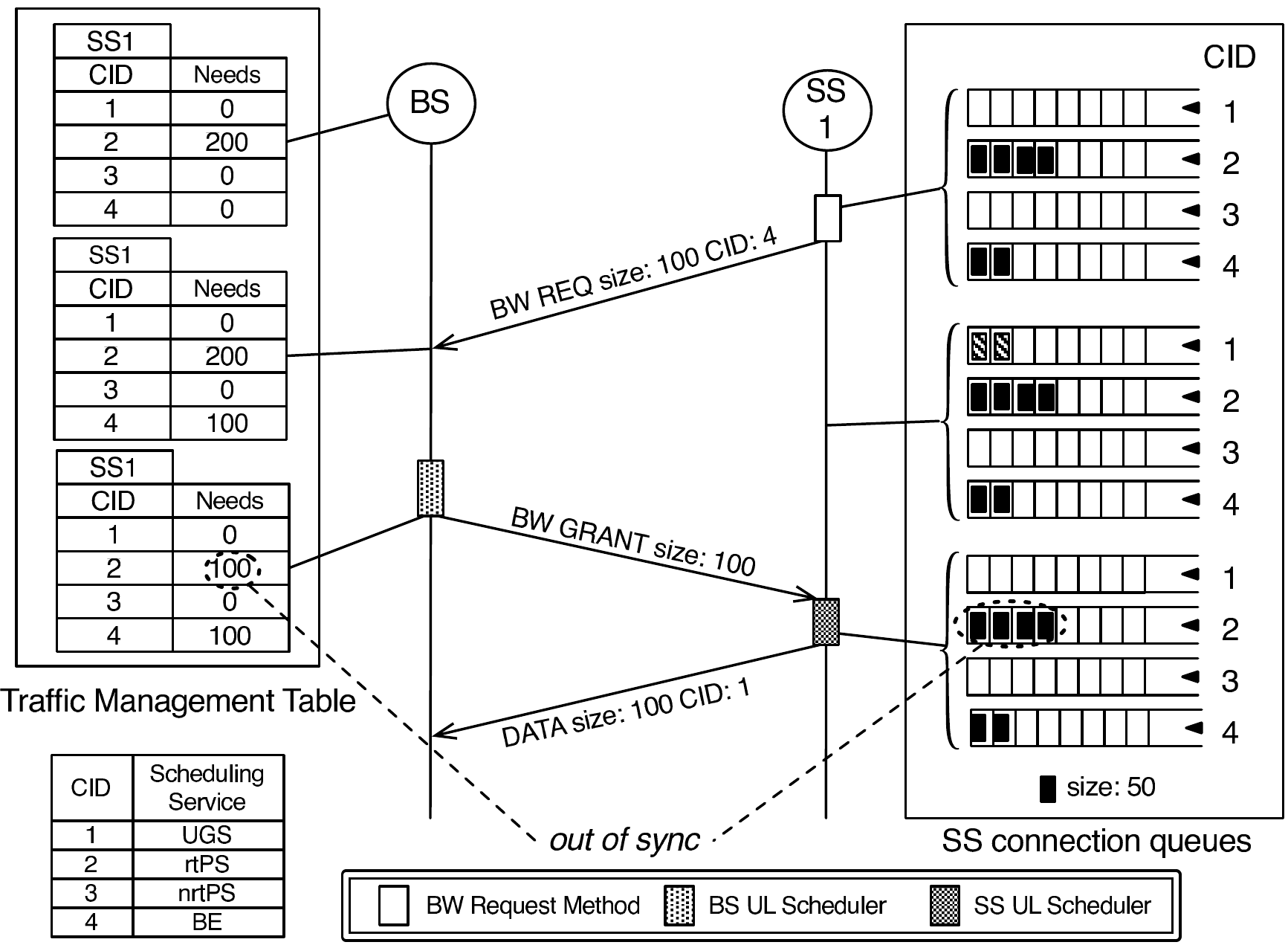}
\caption{DPG (first implementation approach): the BS gets out of sync.}
\label{fig-dpg-1st-approach}
\end{figure}

When new higher-priority packets are queued in the middle of a request-grant exchange (see Fig.~\ref{fig-dpg-1st-approach}), this approach can lead to the BS's perception of the SSs queue status deviating from the actual values: the SS scheduler will serve these higher-QoS packets first, but at this time the BPM will have already decreased the SS's bandwidth needs perception, without taking into account the new packets.

Another approach for the BPM could be as follows: all bandwidth needs of individual connections are grouped per SS, irrespective of the QoS level of connections, and it is this total per-SS amount that is decreased at the time the bandwidth is allocated. The main drawback with this approach is  the loss of the information on the needs of individual connections. Depending on the SS scheduler used, this information loss could be irrelevant. Note that Delicado et~al. \cite{ts:delicado.j;delicado.f08a} propose sending only one request per SS, containing the total amount of resources needed by all its connections, with the goal of decreasing the probability of collision during contention by generating less BW-REQ.

\subsection{Decrease at Data Arrival with immediate BW-REQ handling (DDA-i)}
\label{sec-idda}

With any of the two previous policies, the BS gets out of sync when the SS uses the allocated bandwidth for a purpose different from that originally requested; in the end, how such bandwidth is used is determined by the SS scheduler. In order to correct this deviation, the SS should issue aggregated requests after using the granted bandwidth for a different purpose.

\begin{figure}[tbp]
\centering
\includegraphics[width=0.95\columnwidth]{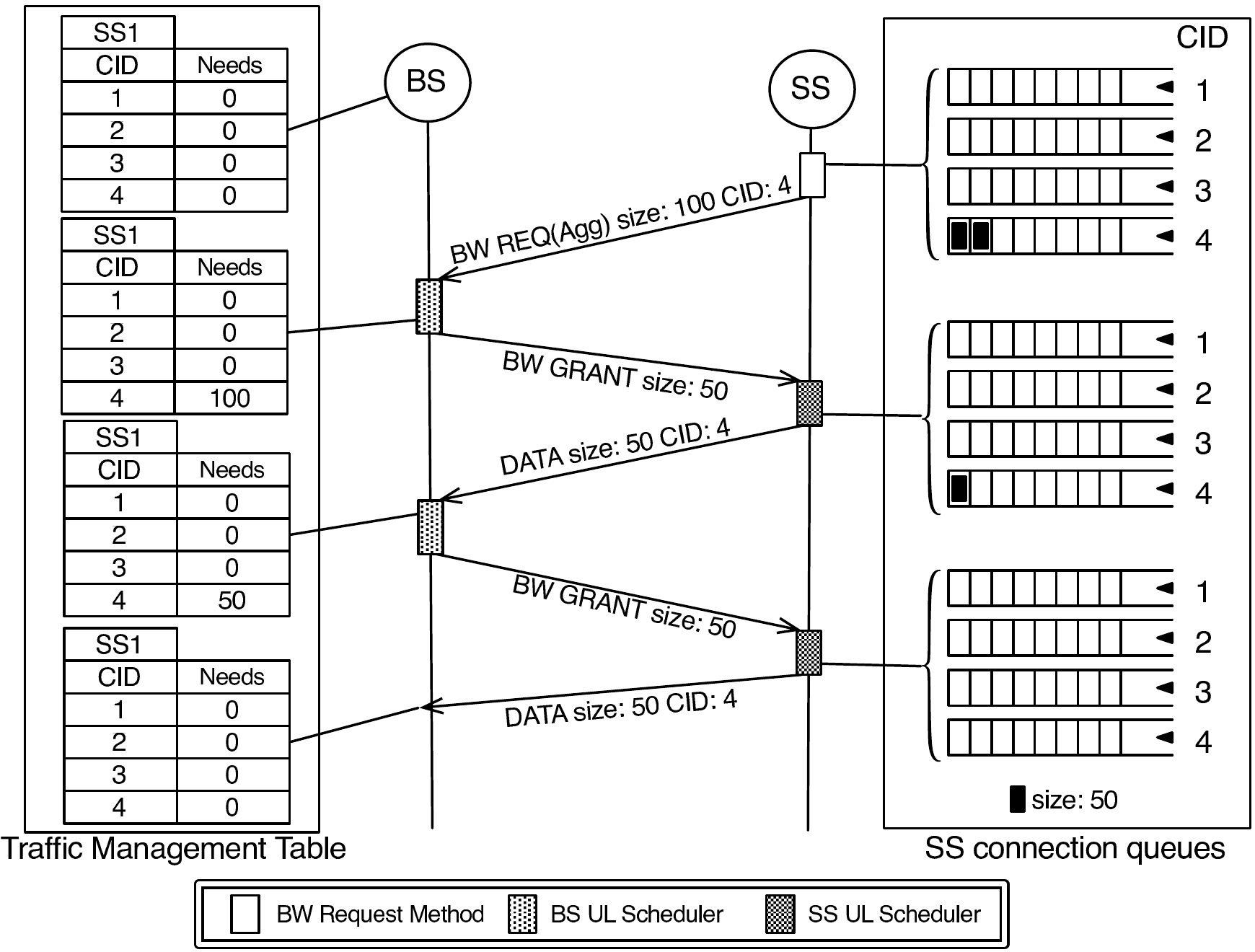}
\caption{DDA with immediate BW-REQ handling (DDA-i) policy.}
\label{fig-iDDA}
\end{figure}

The DDA-i policy solves such problem, by decreasing the perception upon the arrival of successfully transferred \emph{data} frames from the SS to the BS. Therefore, the BPM takes advantage of the CID of incoming data frames in order to \emph{decrease} the perception per connection. BW-REQ messages are, as before, processed as soon as they are received at the BS, so the allocation table is updated at request arrivals. This policy, implemented in the NIST ns-2 simulation model \cite{:rouil.ra}, is exemplified in Fig.~\ref{fig-iDDA}.

This policy could lead to the following undesirable behavior. After receiving a bandwidth grant, the SS scheduler decides which data queue will be served by the granted allocation; then, the SS may ask for more bandwidth if needed, by means of one of the methods described in Section~\ref{sec-uplink-bw-alloc}. Assume that such request is performed using the contention-based method. Since the bandwidth-request contention slots precede the uplink bursts (see Fig.~\ref{fig-wimax-frame}), the BS would receive the new BW-REQ \emph{before} the SS transmits the data. Hence, the allocation table would get out of sync with respect to the actual SS bandwidth needs.

\begin{figure}[tbp]
\centering
\includegraphics[width=0.95\columnwidth]{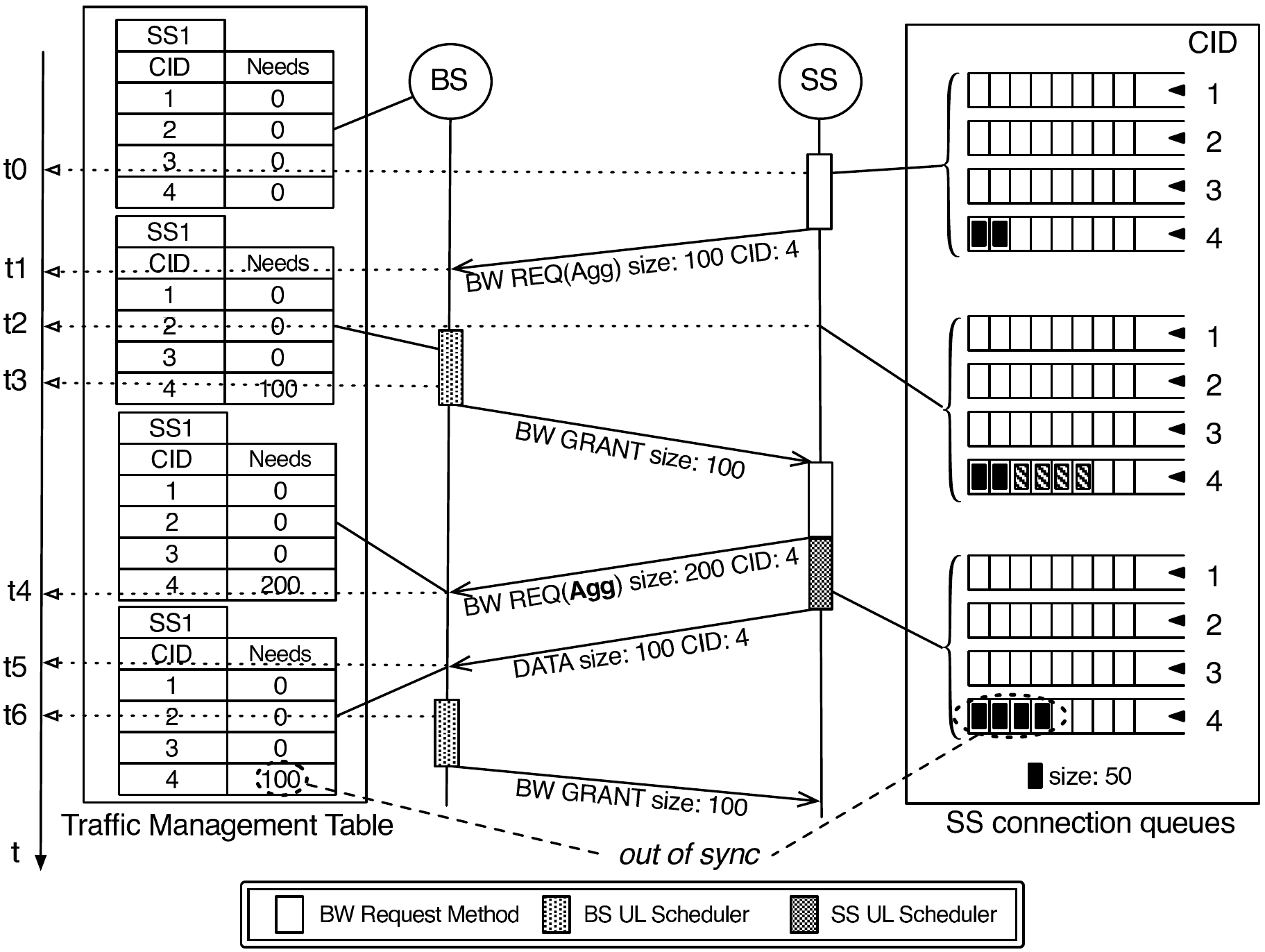}
\caption{DDA-i: the BS gets out of sync.}
\label{fig-iDDA-undesirable}
\end{figure}

Such a problem is illustrated in Fig.~\ref{fig-iDDA-undesirable}. The SS asks for bandwidth using aggregate contention-based requests; assume further that there are no collisions. At $t = t0$, the SS has 100~bytes of data ready to send. At $t = t1$, the BS receives the BW-REQ and immediately updates the allocation table. The BS uplink scheduler grants the bandwidth in the UL-MAP at $t = t3$, allowing the SS to transmit at $t = t5$. Between the request ($t = t1$) and the grant ($t = t3$), 200~bytes of new data are ready to be transmitted ($t = t2$). During the bandwidth contention window of the same TDD frame, the SS asks for more bandwidth taking into account the grant just received at $t=t3$\footnote{Remark that this particular case considers using an aggregate BW-REQ.}. The BW-REQ safely reaches the BS at $t = t4$ and therefore the table is updated with the new requested value of 200~bytes. Later, when the SS transmits the 100~bytes using the granted transmission opportunity, the table is decreased by the transmission size. But, as the table was updated with the BW-REQ, it gets out of sync: the perception of the bandwidth needs is now 100~bytes, instead of the (newly queued) 200~bytes, so the SS will have to resend a request for the remaining 100~bytes.

\begin{figure}[tbp]
\centering
\includegraphics[width=0.95\columnwidth]{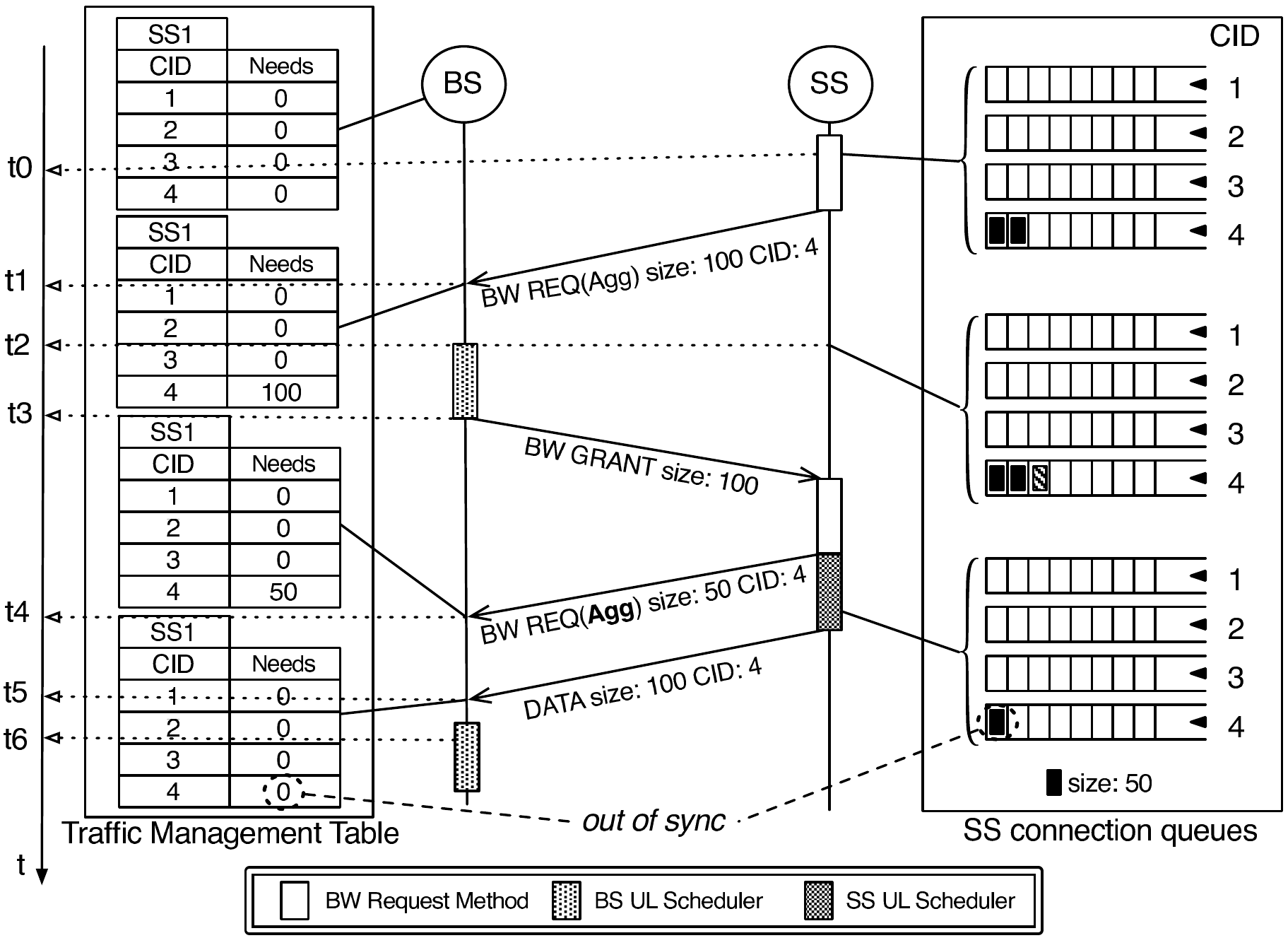}
\caption{DDA-i: worst-case scenario of desynchronization.}
\label{fig-iDDA-worst-case}
\end{figure}

Such undesirable behavior can be worse if the new request size is \emph{smaller} than the current bandwidth perception at the BS, as shown in Fig.~\ref{fig-iDDA-worst-case}. In this case, the perception will be set to zero when the first 100~bytes of data arrive: data flow will be interrupted, the reservation timeout will expire, then the SS will execute the BEB algorithm to request bandwidth again.

\subsection{Decrease at Data Arrival with delayed BW-REQ handling (DDA-d)}
\label{sec-ddda}

The above issues of the DDA-i policy can be fixed by simply delaying the handling of the BW-REQ. That is, the allocation table is updated with the request information just before the execution of the BS uplink scheduler. We call such a policy \emph{Decrease at Data Arrival with delayed BW-REQ handling} (DDA-d).

\begin{figure}[tbp]
\centering
\includegraphics[width=0.95\columnwidth]{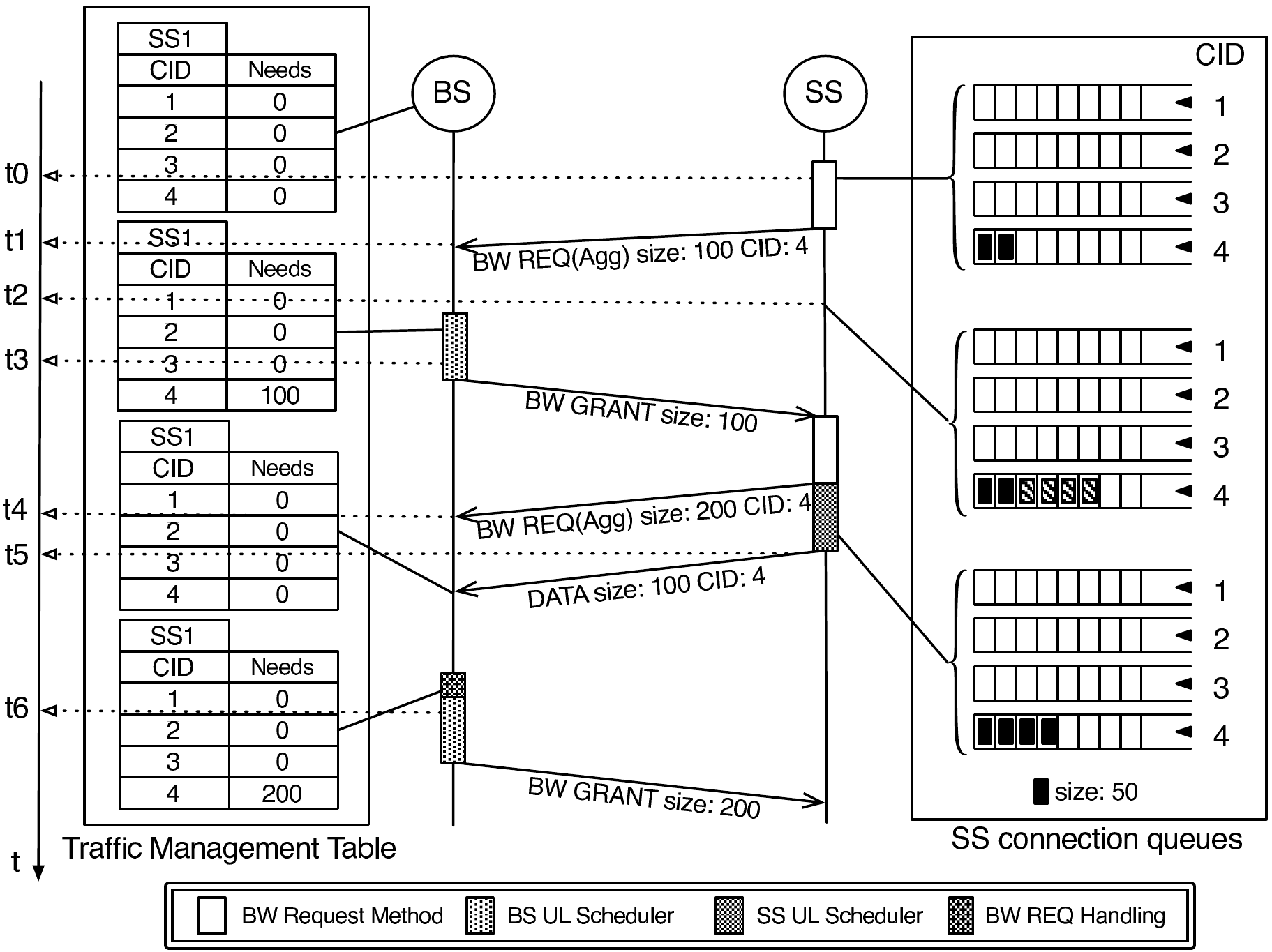}
\caption{DDA-d: with delayed BW-REQ handling (DDA-d) policy.}
\label{fig-dDDA}
\end{figure}

Figure~\ref{fig-dDDA} illustrates DDA-d with the same scenario used in Fig.~\ref{fig-iDDA-undesirable} for the DDA-i policy. Remark that at $t = t4$, when the request reaches the BS, the table is not updated (i.e., the BW-REQ handling is delayed until just before the scheduler execution at $t = t6$), so the table does not lose its consistency.

\section{Performance evaluation}
\label{sec-perfeval}

In this section we present an evaluation of the different perception-management policies discussed before. Our main purpose is not to be exhaustive, but to illustrate how the policies may be sensitive to different scenario configurations.

\subsection{Simulation setup}
\label{sec-simu}

To test the different policies, we extended the WiMAX model developed at the National Institute of Standards and Technology (NIST) \cite{:rouil.ra}, for the well-known \emph{ns-2} simulator.

For this study we set up a number of TCP long-lived flows in the downlink and uplink directions. We independently considered each direction to observe the impact of the uplink bandwidth perception on data packets and ACKs. We tested different numbers of TCP flows, randomly varying the starting time of each flow to avoid synchronization.

\begin{figure}[tbp]
\centering
\includegraphics[width=0.8\columnwidth]{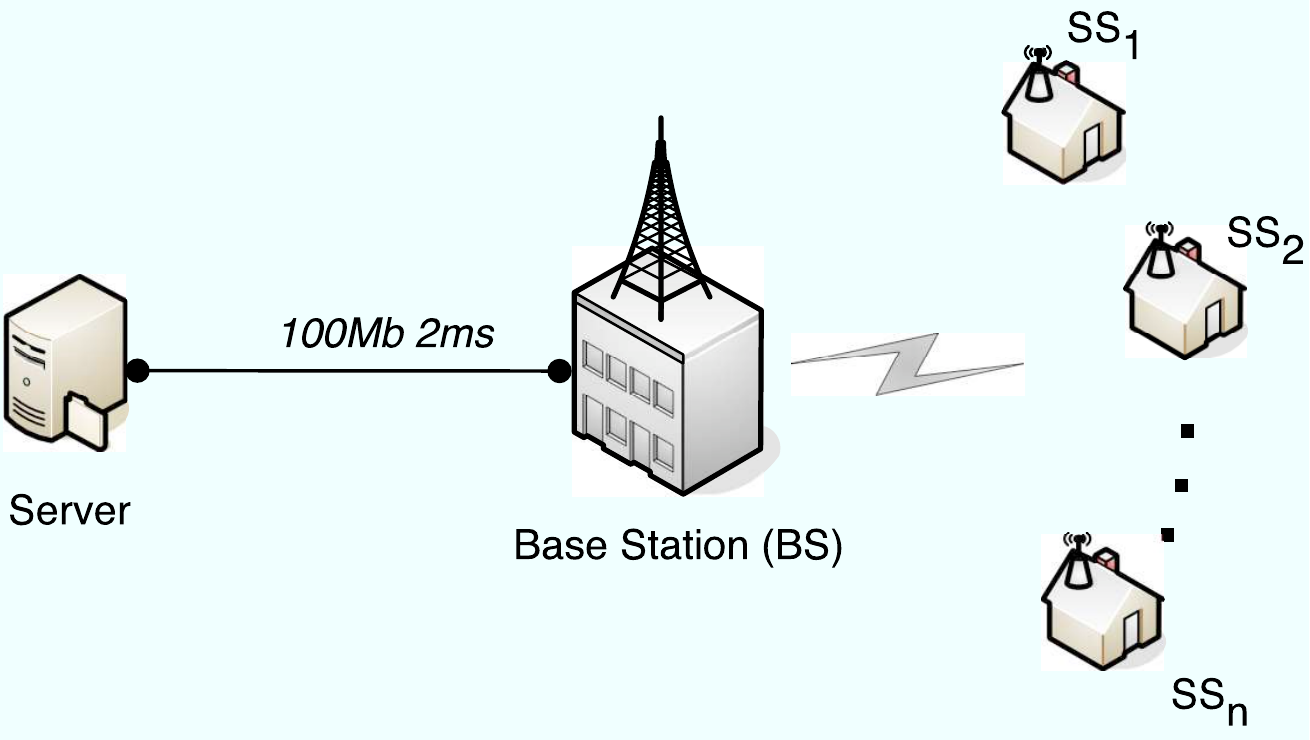}
\caption{Simulation topology.}
\label{fig-topology}
\end{figure}

Figure \ref{fig-topology} shows the system model used for experimentation. Our results hold for 
a WirelessMAN-OFDM physical layer and Time Division Duplexing (TDD) mode to access the media. Unless stated otherwise, we consider ideal channel conditions. The BS scheduler dispatches data packets in a round robin (RR) fashion, as follows. For $N$ connections in the downlink, whenever there is data within a chosen CID queue $i$, the BS fills the downlink subframe as much as possible with data from connection $i$. If there is not enough space within the subframe (i.e., if there remains data from CID $i$ to be sent), the selected SS has to wait a whole round to be served again in the downlink. This same strategy applies for subsequent SSs if one or more SSs are completely served, and there remains free space in the current subframe for others to be served. In Table~\ref{tab-sim-params} we summarize the rest of the simulation setup.

We focused on the following performance metrics:

\begin{itemize}
        \item {\it Aggregated Throughput} corresponds to the sum of individual data throughputs for all unidirectional or bidirectional flows.
        \item {\it Collision Probability} corresponds to the probability that two or more BW-REQs collision during a contention period. Recall that collision resolutions are handled by the BEB mechanism. 
        \item {\it T16 expiration rate per SS} corresponds to the mean number of expirations of the T16 timer, per SS and per second.
\end{itemize}

\subsection{Results}
\label{sec-results}

\subsubsection{Download-only Traffic Scenario}
\label{sec-download-scenario}

\begin{figure*}[htbp]
\centering
\subfloat[Aggregated Throughput]
{%
\label{fig-idda-throughput-only-aggr}%
\includegraphics[width=0.6\columnwidth]%
{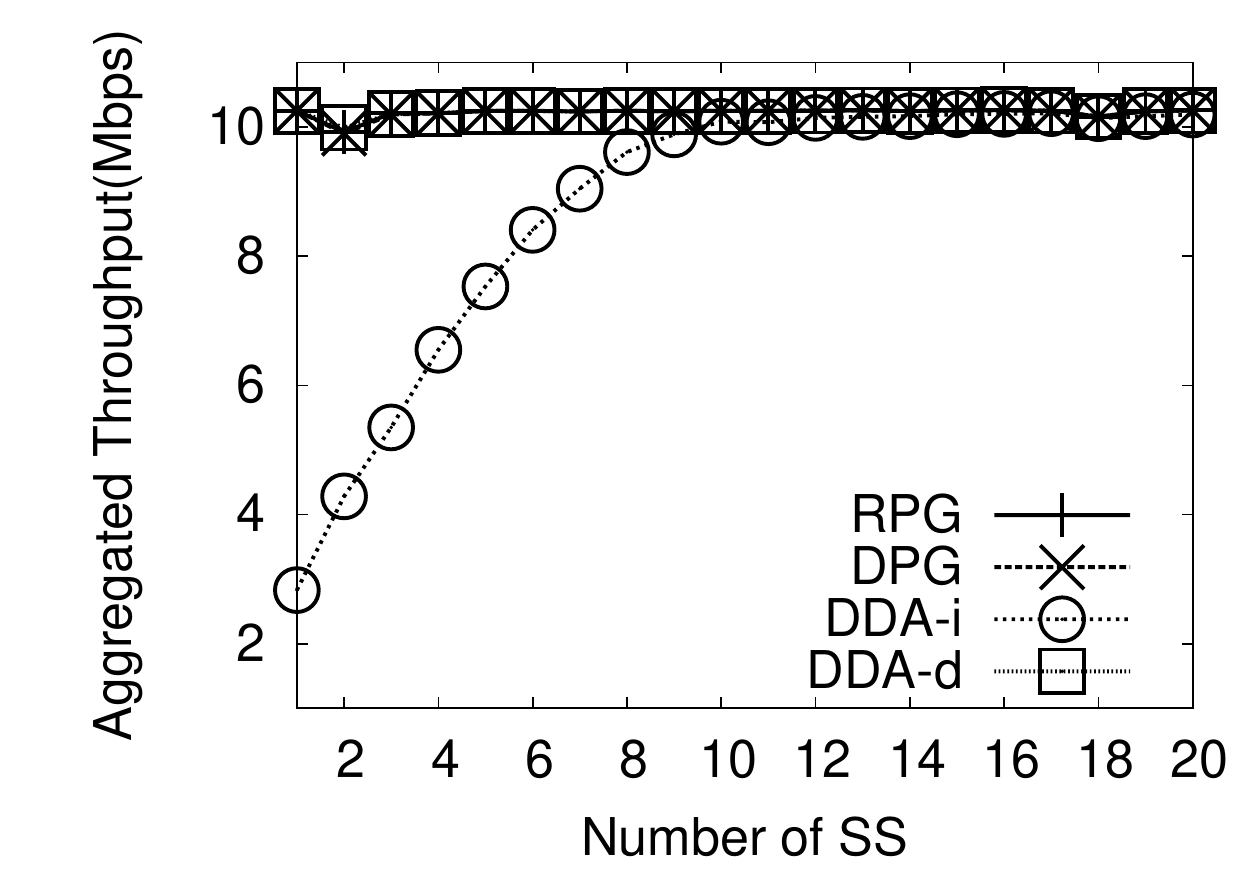}}
\qquad
\subfloat[BW-REQ Collision Probability]
{%
\label{fig-idda-collision-probability}%
\includegraphics[width=0.6\columnwidth]%
{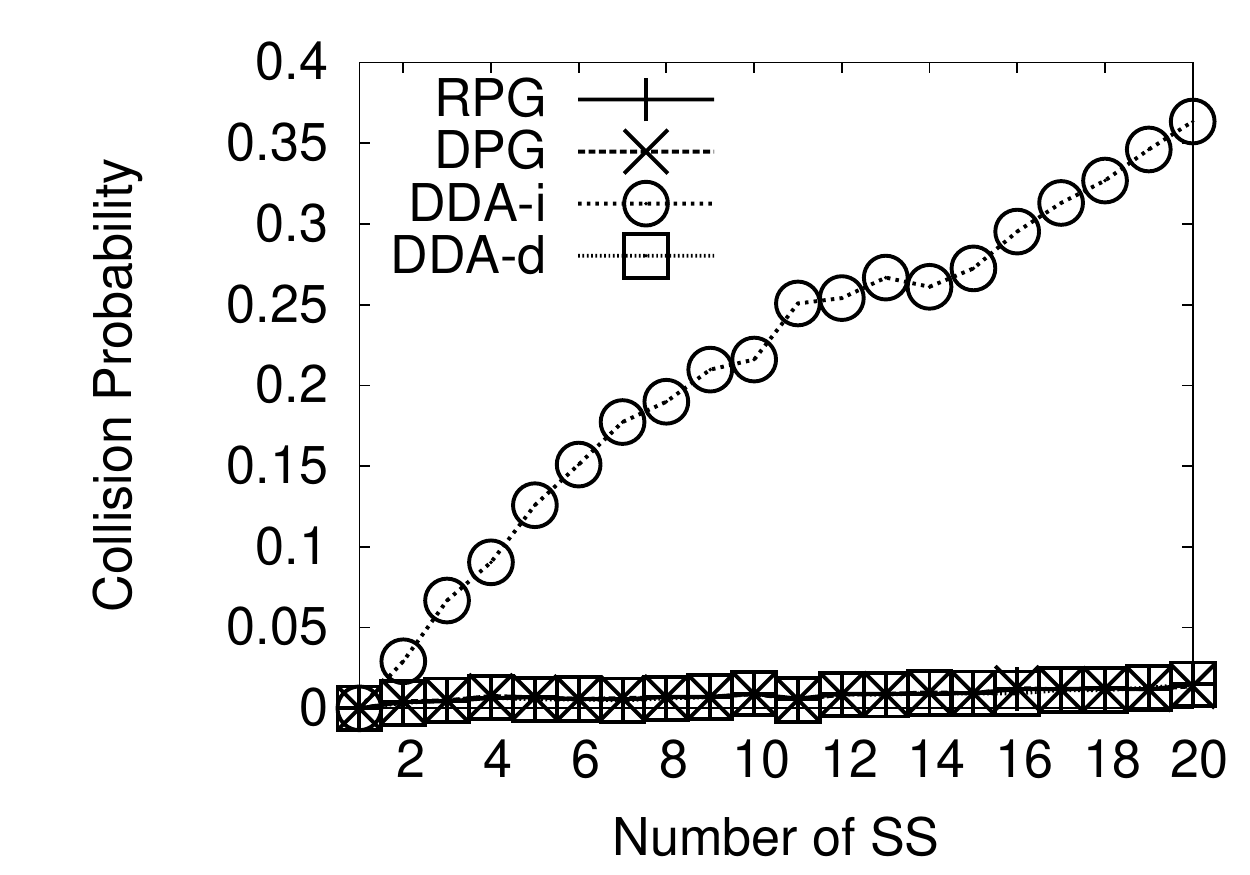}}
\qquad
\subfloat[T16 Expiration Rate per SS]
{%
\label{fig-idda-t16-expiration-rate}%
\includegraphics[width=0.6\columnwidth]%
{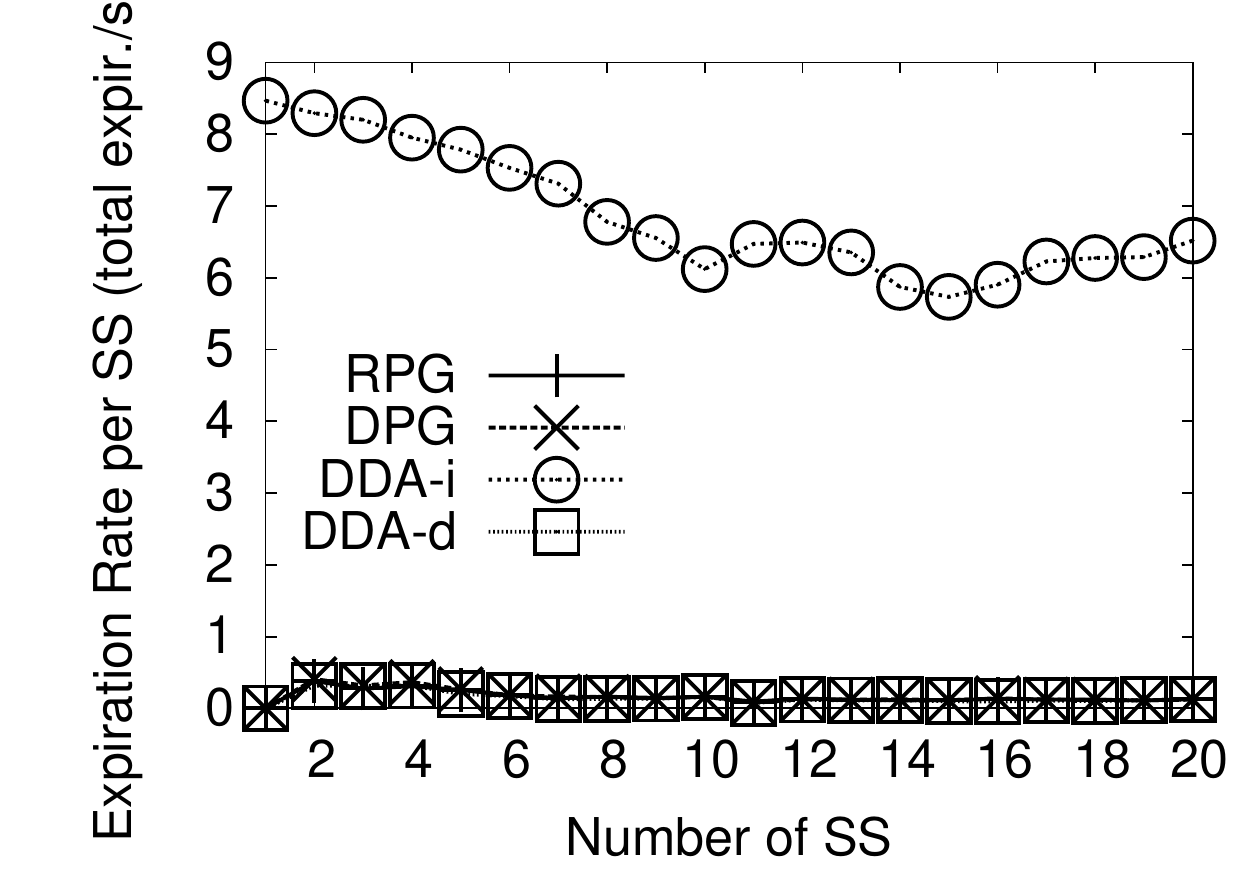}}
\caption{Download-only traffic scenario, using exclusively aggregate requests (BW-REQs) for asking for uplink bandwidth.}
\label{fig-bg-impact-b-to-a}
\end{figure*}

In this scenario we increase the number of SSs; each SS uses a single TCP long-lived flow for downloading data from the server. %

As shown in Fig. \ref{fig-idda-throughput-only-aggr}, DDA-i is the only policy to present a significant underutilization of the downlink bandwidth for a small number of SSs. The rest of the policies perform similarly better than DDA-i. The explanation for the degraded performance is the %
loss of synchronization proper to the \mbox{DDA-i} policy. Actually, as a result of the bandwidth underestimation the DDA-i policy interrupts the (uplink) TCP ACK traffic, inducing repetitive timeouts. This phenomenon is also observed through the increased BW-REQ collision probability in Fig.~\ref{fig-idda-collision-probability} and T16 expiration rate in Fig.~\ref{fig-idda-t16-expiration-rate}, which are notably higher for DDA-i than for the rest of the policies. On the other hand, the aggregated throughput reaches the maximum for the DDA-i policy as the number of SS increases. This happens because, when some uplink flows are ``interrupted'', some other stations have packets in downloading CID queues ready to be dispatched---recall that the BS uplink scheduler can fill a whole frame with data from other (uninterrupted) flows. Finally, the rest of the policies guarantee a low collision and T16 expiration rate in this scenario.

\begin{table}
\renewcommand{\arraystretch}{1.3}
\begin{threeparttable}
\caption{Simulation parameters}
\label{tab-sim-params}
\centering
\begin{tabular}{l|l}
\hline
Parameter  &  Value \\
\hline 
\hline 
\emph{802.16 parameters:} & \\
Carrier frequency  & 3.5 GHz \\
Channel bandwidth  & 7 MHz \\
Frame duration  & 5 ms \\
Modulation and coding  & 64 QAM 3/4 \\
DL:UL ratio & 0.5 \\
T16 timer\tnote{a} & 100 ms \\
\emph{Traffic and other parameters:} & \\
Application type & Long-lived file transfer \\
TCP version & Newreno \\
TCP segment size (without headers) & 960 bytes \\
TCP ACK scheme & Delayed ACKs \\
BS queue type & Drop-Tail \\
BS default queue size & 50 packets \\
SS default queue size & 50 packets \\
Simulation duration & 1000 seconds \\
\hline 
\end{tabular}
\begin{tablenotes}
\item [a] A faster timer could get triggered more frequently, increasing further the BW-REQ collisions for DDA-i. Thus, we chose a fairly long T16 value (with respect to the frame duration) so as to lessen the impact of such effect on the bandwidth perception. A detailed study of what is an ``adequate'' value for T16 is left as future work.
\end{tablenotes}
\end{threeparttable}
\end{table}

\subsubsection{Upload-only Traffic Scenario}

In this scenario we test an increasing number of SSs uploading data. We set a single TCP connection from each of the SSs to the server. For this scenario we emphasize the impact of mixing aggregated and incremental BW-REQs\footnote{For space reasons, we do not show the results for the scenario with aggregated BW-REQs only; in such case, all the proposed policies produce the maximum throughput with a fair share of the bandwidth.}. Note that, due to the use of delayed ACKs by TCP receivers, during the analyzed phase (i.e., TCP congestion avoidance) every incoming ACK at a TCP sender triggers the transmission of two data packets. 

\begin{figure*}[htbp]
\centering
\subfloat[Aggregated Throughput]
{%
\label{fig-rpg-throughput-mixed}%
\includegraphics[width=0.6\columnwidth]%
{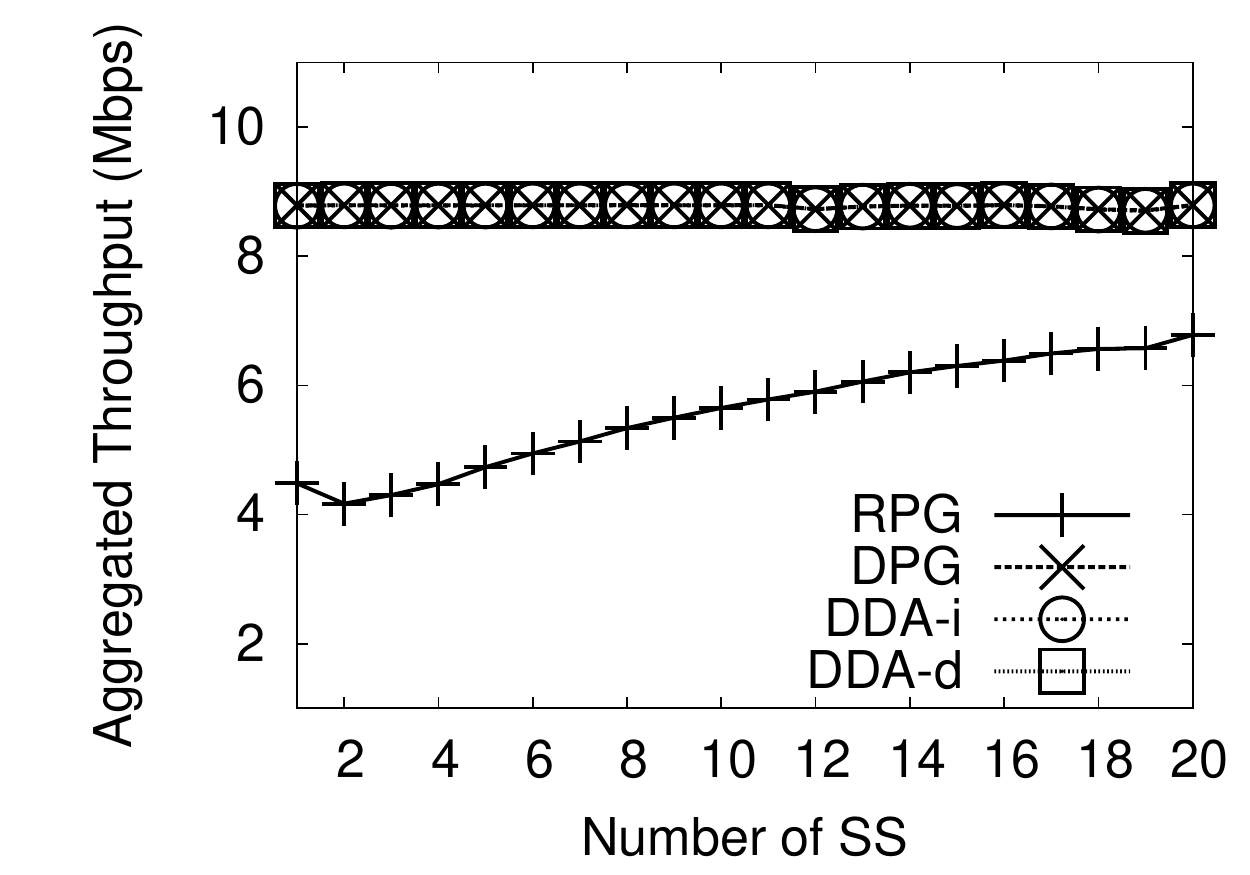}}
\qquad
\subfloat[BW-REQ Collision Probability]
{%
\label{fig-rpg-collision-prob-mixed}%
\includegraphics[width=0.6\columnwidth]%
{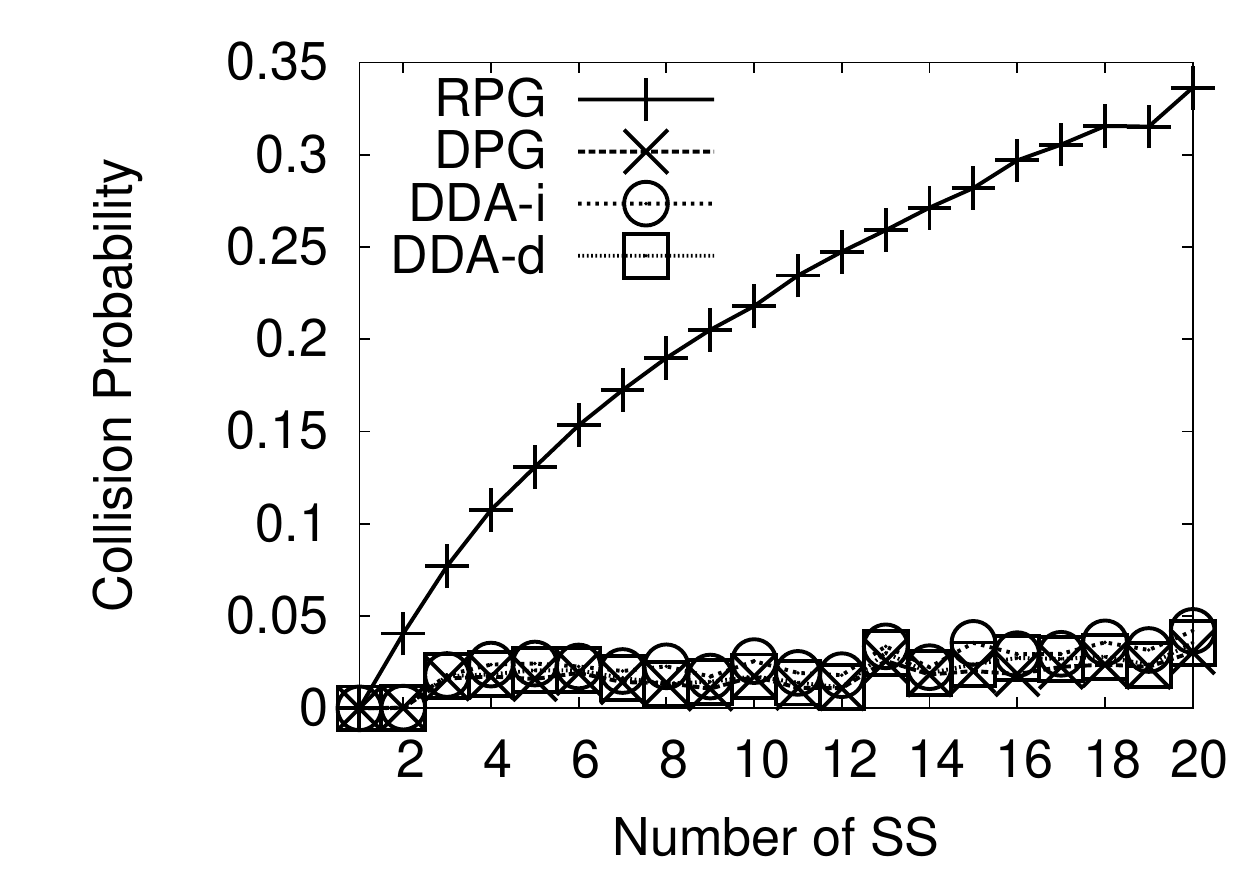}}
\qquad
\subfloat[T16 Expiration Rate per SS]
{%
\label{fig-rpg-t16-exp-rate-mixed}%
\includegraphics[width=0.6\columnwidth]%
{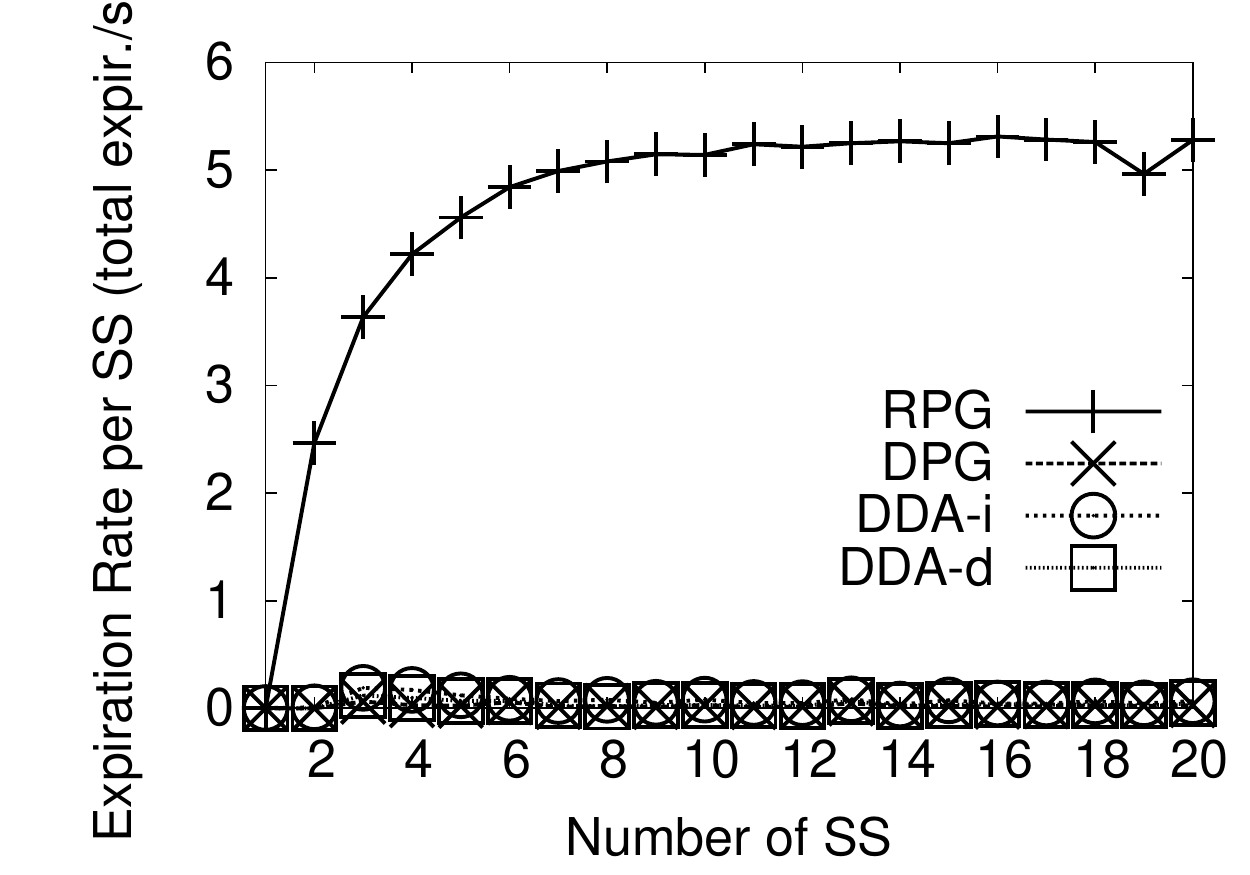}}
\caption{Upload-only traffic scenario, using one aggregate bandwidth request per 50 incremental bandwidth requests (BW-REQs).}
\label{fig-upload-traffic}
\end{figure*}

To illustrate the effect of incremental BW-REQs, we send one aggregated BW-REQ for every 50 incremental BW-REQs. Since at the BW-REQ arrival the scheduler either grants part of the request or all the underestimated solicited bandwidth (i.e., smaller grants than actual needs), most of the time the BS will misperceive the real needs of the SSs. Consequently, the RPG policy degrades the aggregated throughput of the upload traffic, as seen in Fig.~\ref{fig-rpg-throughput-mixed}. As Fig.~\ref{fig-rpg-collision-prob-mixed} shows, in most of the frames the SSs have to generate BW-REQs due to the constant backlog in the uplink queues, hence the high number of collisions. Finally, Fig.~\ref{fig-rpg-t16-exp-rate-mixed} shows that the high BW-REQ collision rate can easily induce T16 expirations, deteriorating even further the upload throughput when using the RPG policy.

\subsubsection{Impact of the BS queue length}
In this scenario, we show the impact of MAC-layer queue length limits. Each SS downloads data from the server. Since downlink data packets are queued at the BS, the size of the SS queue size (which only holds TCP ACKs) is set to 20 packets.

\begin{figure*}[htbp]
\centering
  \subfloat[Aggregated Throughput.]
  {%
  \label{fig-queue-throughput}%
  \includegraphics[width=0.6\columnwidth]%
  {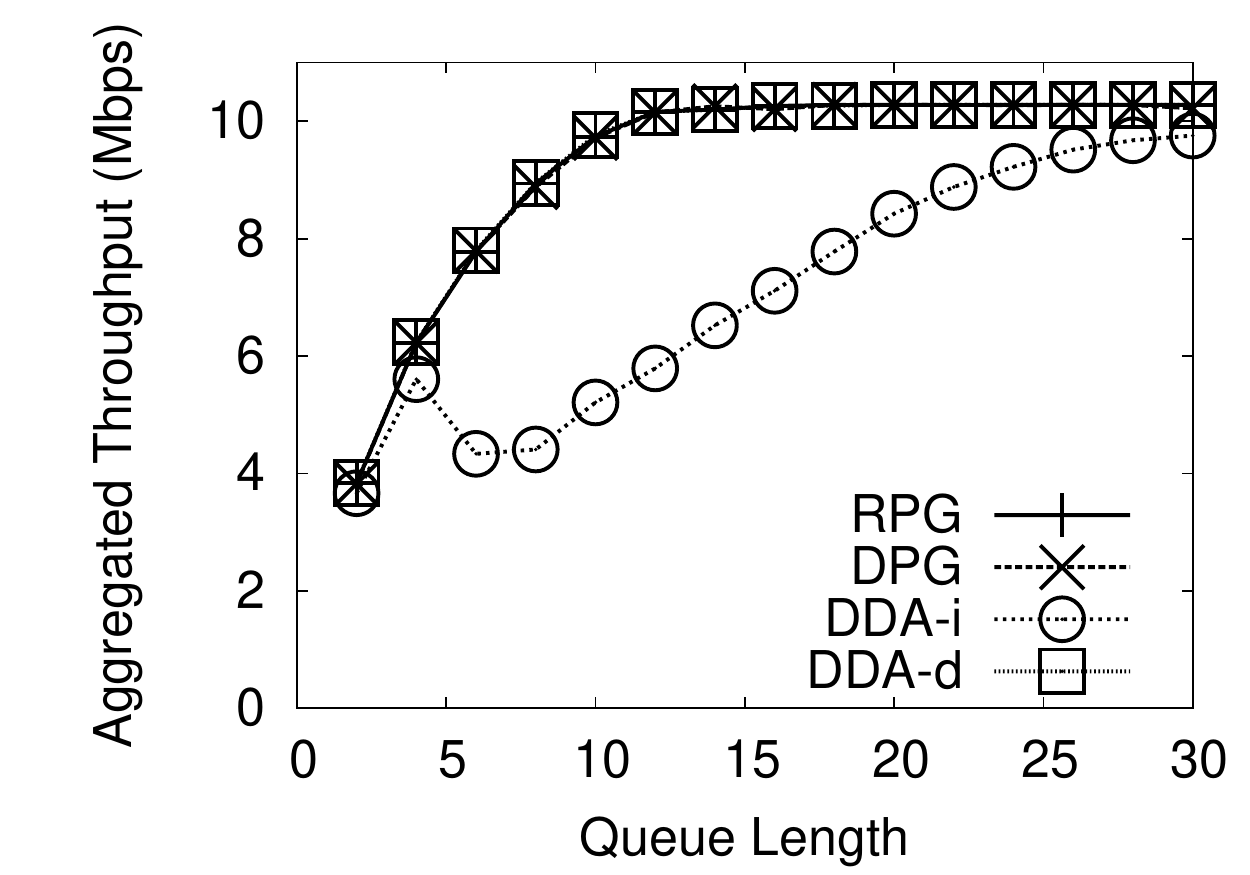}}
\qquad
  \subfloat[BW-REQ Collision Probability.]
  {%
  \label{fig-queue-collision}%
  \includegraphics[width=0.6\columnwidth]%
  {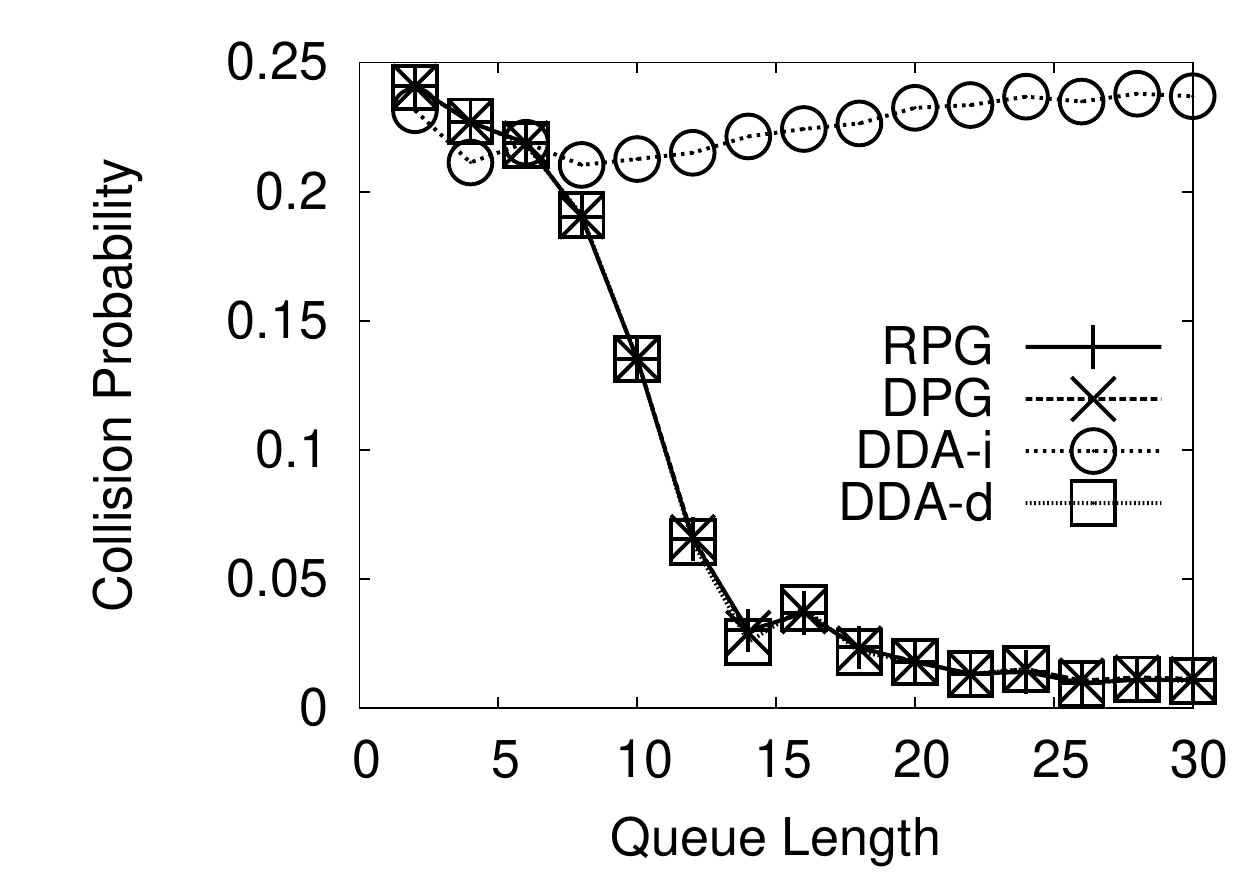}}
\qquad
  \subfloat[T16 Expiration Rate per SS.]
  {%
  \label{fig-queue-t16-exp-rate}%
  \includegraphics[width=0.6\columnwidth]%
  {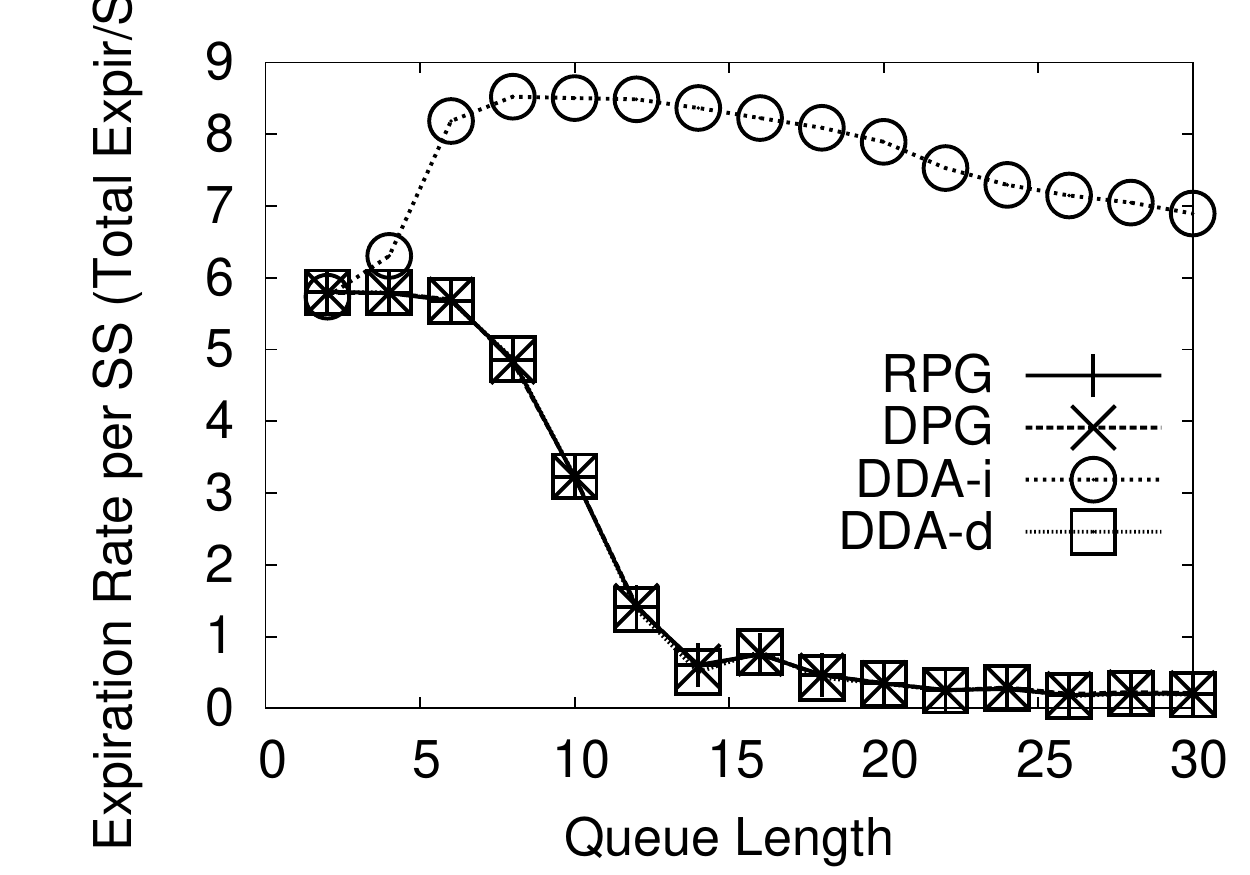}}
\caption{Downlink only traffic, Performance with different queue length limits. The number of SSs is 10.}
\label{fig-queue-performance}
\end{figure*}

Figure~\ref{fig-queue-performance} gives the aggregated throughput, BW-REQ collision probability and T16 expiration rate with different BS queue sizes. From Fig.\ \ref{fig-queue-throughput}, we see that total throughput increases as the BS queue size increases. This is because the base station acts as a bottleneck in our scenario, and thus the BS queue has an impact on the TCP congestion window size at server. When queue sizes are small, since TCP data packets are sent in a bursty manner, then queue overflow will occur if the congestion window is larger than the queue length limit, and thus packet drop occurs. Therefore, the BS queue size constrains the server's TCP congestion window size. When queue sizes are large, queuing delay increases and it is the transmitting rate over the wireless link that constrains TCP throughput. 

Note that the aggregated throughput for RPG, DPG and DDA-d reaches a maximum, stable value for BS queue sizes of more than 10 packets. On the other hand, total throughput for DDA-i is lower; DDA-i requires larger queues in order to reach the maximum throughput.
For DPG, RPG and DDA-d, as the queue lengh limit increases, more data packets are received by a SS in a burst, and thus one BW-REQ serves more ACK packets. Therefore, as queue length limit increases, less BW-REQ are transmitted, and thus collision probability decreases, which is shown in Fig. \ref{fig-queue-collision}. Similarly, T16 expiration rate also decreases for RGP, DPG and DDA-d, which is depicted by Fig. \ref{fig-queue-t16-exp-rate}.
However, for DDA-i, its collision probability and T16 expiration rate do not decrease as queue size increases. This is because BW-REQ transmissions with DDA-i are strongly driven by the desynchronization of bandwidth perception, which does not improve noticeably with increasing queue sizes. %

\subsubsection{Impact of wireless losses}

In this section, we study the impact of random wireless loss on the performance of the four policies. In this scenario, each SS downloads data from the server. The wireless loss rate, in terms of packet error rate, is simulated by a Gilbert-Elliot error model. The model is a two-state Markov chain including ``good'' and ``bad'' channel states, with different packet loss rates $P_g$ and $P_b$, respectively. In addition, the channel state changes from good to bad with probability $P_{gb}$, and from bad to good with probability $P_{bg}$. From such values, we compute the steady-state probabilities of good and bad channel states $\pi_g$ and $\pi_b$. In the simulation, we fix $P_g$, $P_{gb}$ and $P_{bg}$, and change $P_b$ to get different average packet loss rates $P_{loss}=\pi_g P_g + \pi_b P_b$. Numerical values used were: $P_g = 0$, $P_{gb} = P_{bg} = 0.5$, so $\pi_g = \pi_b = 0.5$.

\begin{figure*}[htbp]
\centering
  \subfloat[Aggregated Throughput.]
  {%
  \label{fig-wireless-loss-throughput}%
  \includegraphics[width=0.6\columnwidth]%
  {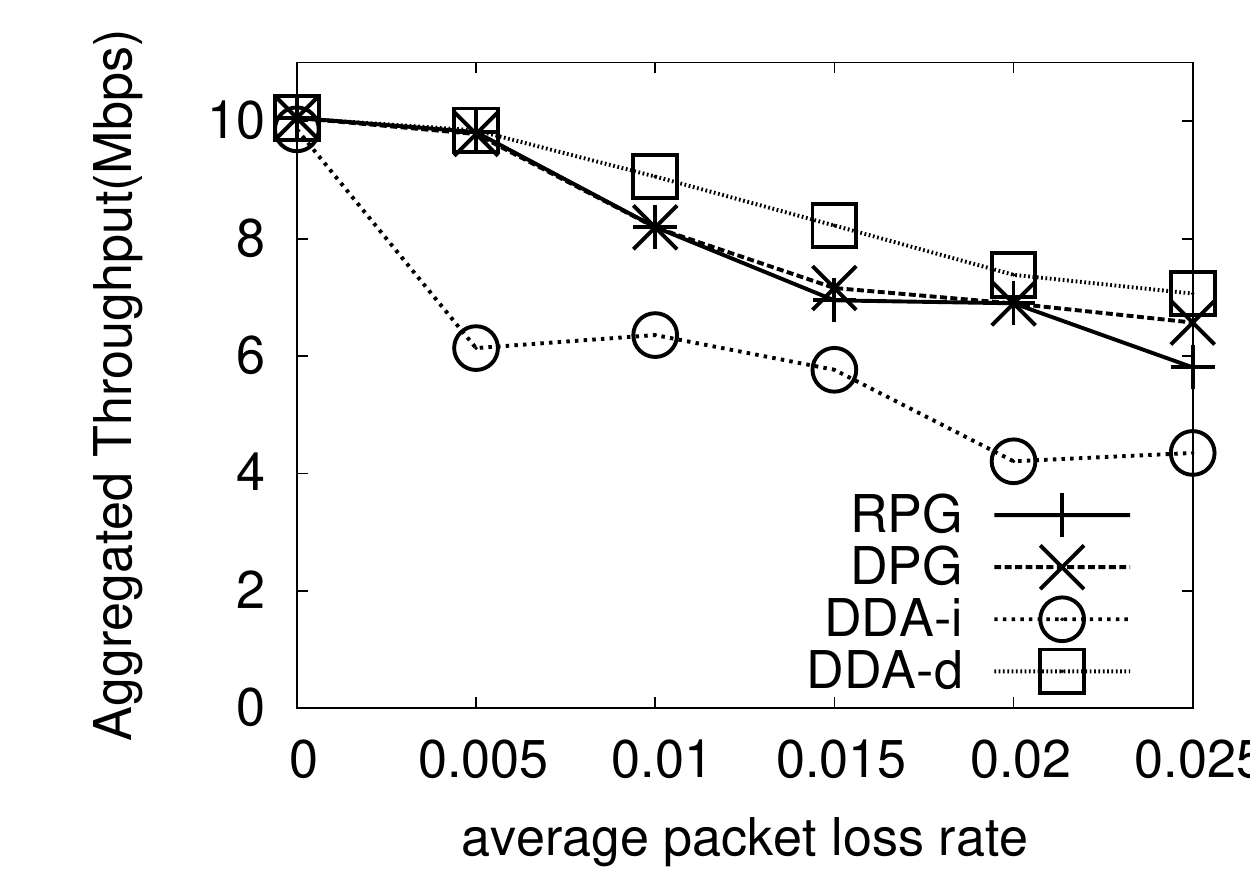}}
\qquad
  \subfloat[BW-REQ Collision Probability.]
  {%
  \label{fig-wireless-loss-collision}%
  \includegraphics[width=0.6\columnwidth]%
  {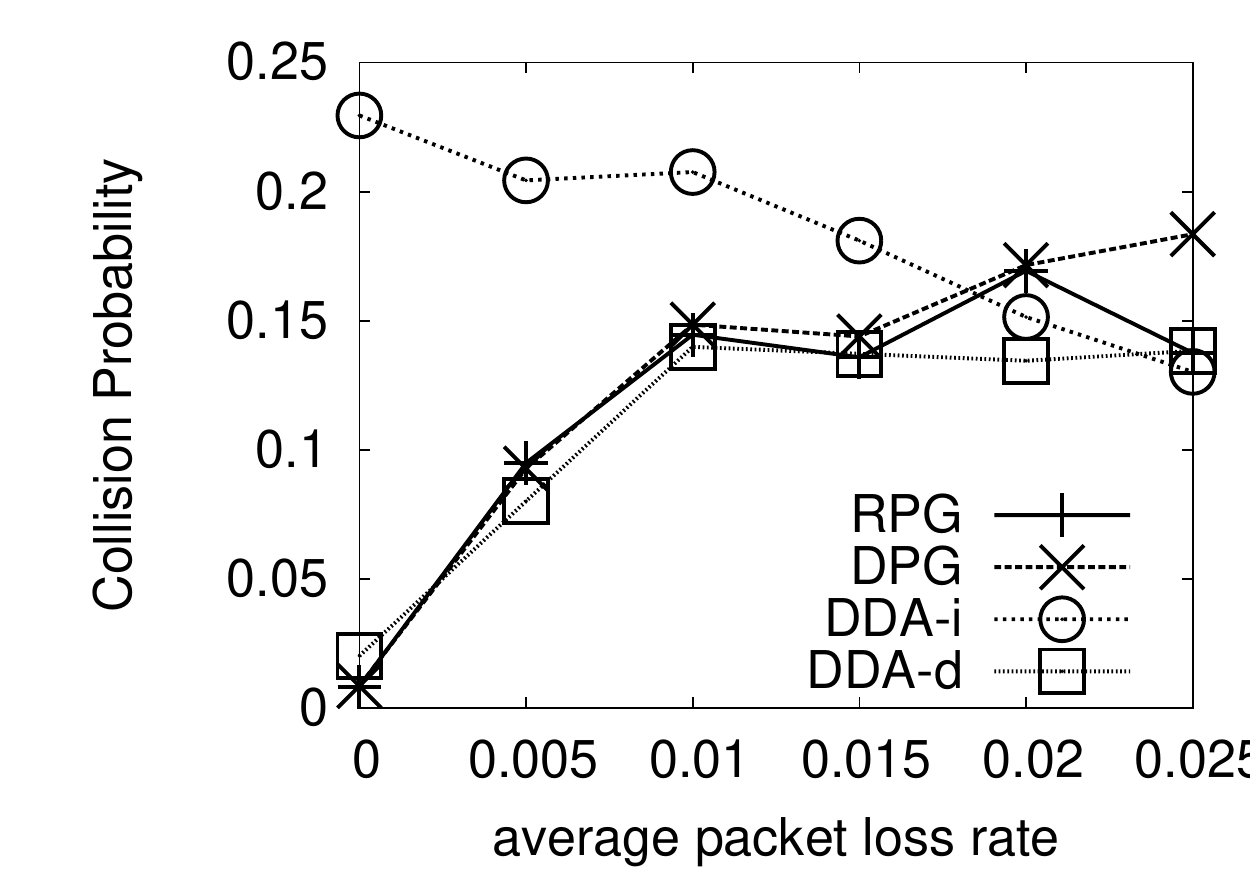}}
\qquad
  \subfloat[T16 Expiration Rate per SS.]
  {%
  \label{fig-wireless-loss-t16-exp-rate}%
  \includegraphics[width=0.6\columnwidth]%
  {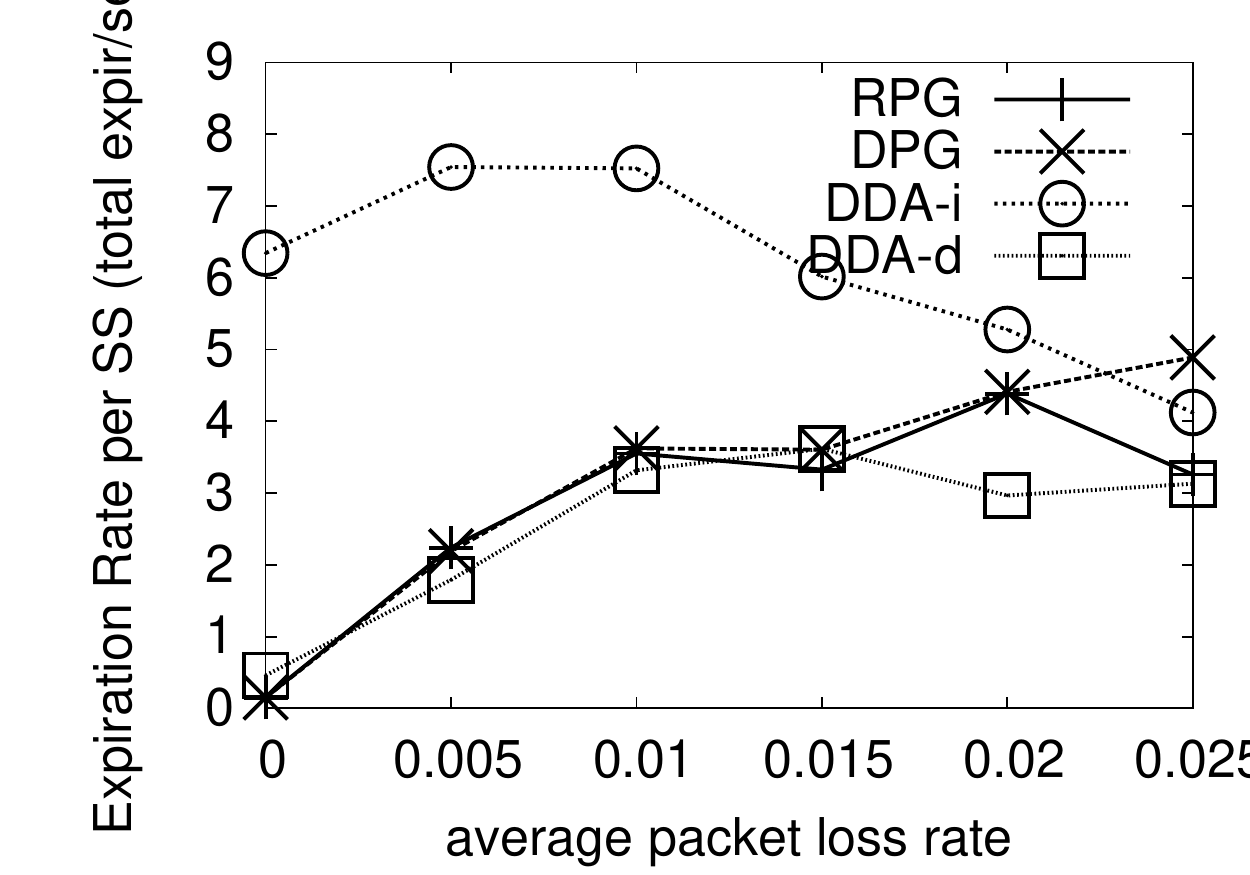}}
\caption{Downlink only traffic, Performance with different wireless loss rates. The number of SSs is 10.}
\label{fig-wireless-loss-performance}
\end{figure*}

From Fig.\ \ref{fig-wireless-loss-throughput}, we can notice that the aggregated throughput decreases as wireless loss rate increases. This is an expected result, as TCP reacts to random losses by (inappropriately) reducing its congestion window. %

For RPG, DPG and DDA-d, since some bandwidth requests are dropped due to wireless losses, more retransmissions of BW-REQ packets are generated. Therefore, the collision probability and the timer expiration rate increase as the wireless loss rate gets higher  (Figs.\ \ref{fig-wireless-loss-collision} and \ref{fig-wireless-loss-t16-exp-rate}). On the other hand, their rate of increase seems to get lower as the wireless loss rate increases. This is because TCP throughput decreases with a higher rate of packet errors, and thus less BW-REQ packets are generated to request for uplink bandwidth. Note that DDA-i behaves quite differently in the presence of random losses. With \mbox{DDA-i}, BW-REQ transmissions are mainly driven by the ``out-of-sync'' property; thus, as throughput decreases, collision probability and T16 expiration rate decrease as well.

\section{Conclusions and Future Work}
\label{sec-conclusions}

In this paper we have studied different policies for managing the bandwidth perception at the BS. We identified a set of critical cases in which the BS may misperceive the bandwidth needs at the SSs. 
To improve such perception, we have suggested a variant of an existing policy, which we call Decrease at Data Arrival with delayed processing of BW-REQs. 
By means of simulation, we have assessed the impact that the bandwidth perception may have on data throughput. We have observed that the incorrect perception of the bandwidth needs at the BS increases the rate of BW-REQs issued by SSs, which in turn leads to an underutilization of the uplink bandwidth. Finally, we have also studied the impacts of the BS queue size and wireless loss on TCP performance with different bandwidth-perception schemes.

The behavior of the policies under consideration could change when considering traffic from classes of service higher than BE. Indeed, the bandwidth perception may have an impact on the scheduling decisions, which in turn may affect the perception. Hence, as future work we would like to study the relation between the scheduling policy and the perception policy. Also deserving further study is the proportion of aggregate and incremental requests in the presence of random losses, and its impact on performance when a particular perception policy is in use.

\section*{Acknowledgments}

This work was partially supported by the French ANR project WiNEM and the Venezuelan Fund for the Development of Telecommunications (FIDETEL). The authors would like to thank Alejandro Paltrineri for his help with a preliminary version of this work.

\IEEEtriggeratref{28}

\bibliographystyle{IEEEtran}
\bibliography{biblio}

\end{document}